# Designing Patient-Specific Optimal Neurostimulation Patterns for Seizure Suppression

Roman A. Sandler [*1,2], Kunling Geng [3], Dong Song [3], Robert E. Hampson [4], Mark R. Witcher [5], Sam A. Deadwyler [4], Theodore W. Berger [3] & Vasilis Z. Marmarelis[3]

[1]Department of Physics & Astronomy, University of California, Los Angeles, Los Angeles, CA, USA
[2]W. M. Keck Center for Neurophysics, University of California, Los Angeles, Los Angeles, CA, USA
[3]Department of Biomedical Engineering, University of Southern California, Los Angeles, CA, USA
[4]Department of Physiology & Pharmacology, Wake Forest University, Winston-Salem, NC, USA
[5]Department of Neurosurgery, Wake Forest University, Winston-Salem, NC, USA

January 15, 2018

## Abstract

*Neurostimulation is a promising therapy for abating epileptic seizures. However, it is extremely difficult to identify optimal stimulation patterns experimentally. In this study human recordings are used to develop a functional 24 neuron network statistical model of hippocampal connectivity and dynamics. Spontaneous seizure-like activity is induced in-silico in this reconstructed neuronal network. The network is then used as a testbed to design and validate a wide range of neurostimulation patterns. Commonly used periodic trains were not able to permanently abate seizures at any frequency. A simulated annealing global optimization algorithm was then used to identify an optimal stimulation pattern which successfully abated 92% of seizures. Finally, in a fully responsive, or "closed-loop" neurostimulation paradigm, the optimal stimulation successfully prevented the network from entering the seizure state. We propose that the framework presented here for algorithmically identifying patient-specific neurostimulation patterns can greatly increase the efficacy of neurostimulation devices for seizures.*

## 1 Introduction

Epilepsy is a neurological disorder characterized by chronic seizures which affects 1-2% of the US population (Begley et al., 2000). Standard treatments include antiepileptic drugs and resective surgery. However, both have major drawbacks. Up to 30% of patients do not respond to drugs; of those who do, many suffer serious side-effects such as nausea, dizziness, drowsiness, and weight-gain (Brodie and Dichter, 1996). Furthermore, Surgery is not an option for many patients, and when it is, there is a large remission rate within 1-2 years (Engel et al., 2003).

In recent years, neurostimulation has emerged as a promising approach to reducing seizures. In 2013, the FDA approved the Neuropace RNS system, the first device for responsive cortical neurostimulation for epilepsy (Sun, Morrell, and Wharen Jr, 2008). However, thus far results have shown that neurostimulation provides only palliative relief from seizures rather than a full cure. For example, the Neuropace device has provided a 60% decrease in median seizure after three years (Bergey et al., 2015; Morrell and Halpern, 2016). While these results are impressive, especially considering they were obtained on the most difficult patient category, they are far from perfect. For example, 42% of patients did not respond to treatment in the same time period, and no patients were completely seizure free (Bergey et al., 2015).

---

[*]Corresponding Author: rsandler00@gmail.com.

Previous research has shown that the efficacy of neurostimulation could be increased by carefully designing the temporal pattern of stimulation pulses. Frequency of stimulation (Chkhenkeli and Chkhenkeli, 1997; Cordeiro et al., 2013), periodicity (Wyckhuys et al., 2010; Buffel et al., 2014), and, in the case of multiple electrodes, synchronicity (Good et al., 2009; Nelson et al., 2011; Van Nieuwenhuyse et al., 2014) have all been shown to influence the success of neurostimulation. Furthermore, many have argued that stimulation must be custom tailored to the unique seizure topology and dynamics of each particular patient (Holt and Netoff, 2014; Mina et al., 2013; Taylor et al., 2015). Presumably, optimal stimulation delivered at the optimal time via a responsive closed-loop stimulation paradigm could significantly increase therapeutic outcome (Chakravarthy et al., 2009a; Chakravarthy et al., 2007). However, designing optimal neurostimulation patterns is extremely challenging in animal models and in human patients. Researchers cannot order seizures "on-demand" to test a wide range of stimulus patterns. Oftentimes, physicians must wait months before learning if a particular stimulus works. Furthermore, researchers can only stimulate each seizure once, and cannot go "back in time" to see how the seizure would have evolved with no stimulation or with different stimulation. Due to these difficulties the Neuropace device recommends that physicians keep stimulus frequency fixed at 200 Hz and only if that fails to adjust the stimulus current. Nonetheless, there is an increasing feeling in the field that a more principled approach to stimulation design is needed (Nagaraj et al., 2015).

In past work we have developed a closed-loop model of the rodent hippocampus and used this model to identify the optimal frequency of stimulation needed to reduce network output (Sandler et al., 2015a). However, the above model was limited to 2 reciprocally connected neurons and thus important features of epilepsy such as population synchrony could not be studied. Here we use 24 neurons recorded from human hippocampus to reconstruct the patient's distinct neuronal connectivity and causal dynamics. Spontaneous seizure activity is then initiated in the reconstructed neuronal network (RNN). Finally, this model is used as an *in-silico* testbed for designing and testing efficient neurostimulation patterns. The optimal stimulus, obtained via a global optimization algorithm, was found to abate 92% of seizures, and significantly outperformed any other traditional stimulation types. We believe that such a patient-specific algorithmic approach to neurostimulation design can significantly increase the efficacy of neurostimulation devices for epilepsy.

## 2 Methods

### 2.1 Human Data

One adult patient suffering from pharmacologically refractory temporal lobe epilepsy was surgically implanted with FDA-approved hippocampal electrodes capable of field potential (macro-) and single-unit (micro-) recordings (Ad-Tech Medical Instrumentation Corporation, Racine, WI) for localization of seizures. All procedures were reviewed and approved by the Institutional Review Board of Wake Forest Baptist Medical Center, in accordance with National Institutes of Health guidelines. Inclusion in this study was voluntary and consented separately from the surgical procedure. As the primary contribution of this work is meant to be methodological, data from only a single patient was used to demonstrate proof-of-concept (see discussion).

Preoperative planning and intraoperative placement of depth electrodes was performed using a frameless Brainlab Cranial Navigation System (BrainLab North America, Westchester, IL) to plan and guide electrode entry points, electrode trajectories and target points within the CA3 and CA1 subfields of each hippocampus. Electrode localization was confirmed by comparing preoperative and postoperative MRI in order to identify electrode track and positioning of the "macro" electrode sites with respect to hippocampal morphology. Single unit neural activities (i.e., spike trains) were recorded and isolated using the Blackrock Cervello Elite electrophysiological recording system with a raw data acquisition frequency of 30 kHz. Electrodes were explanted after seizures were localized (14 days). 10 minutes of continuous recordings were used to estimate all models. Spikes were detected and sorted offline with a 500-5,000 Hz bandpass filter and discretized using a 2 ms bin.

## 2.2 Dynamic Connectivity Model Structure and Estimation

The aim of the RNN is to use the observed spiking activity to reconstruct in-silico the distinct effective connectivity of the 24 recorded neurons. In other words, for each neuron the RNN attempts to answer which of the other $R-1$ neurons causally influence it, and what is the dynamical nature of that influence. The methods used here are similar to those of Song et al. (2016).

The firing probability of each neuron at time $t$, $\hat{y}(t)$, was determined by its own past spiking activity and the past and present spiking activity of all other $N$ connected CA3 neurons, $\{x_n(t)\}$ within a finite memory of $M = 100$ ms and modeled using a generalized linear model (GLM) with a probit link (see Eq. A2-A4). Each feedforward/feedback filter is either linear or quadratic-nonlinear (2nd order Volterra) (see Eq. A4), as determined by the group regularization algorithm used for model fitting. The feedback component, characterized by the filter $k_{AR}(\tau)$, can be intuitively thought of as the afterhyperpotential (AHP) (Spruston and McBain, 2007) and encapsulates intracellular processes such as the absolute and relative refractory period, slow potassium conductances, and $I_h$ conductances. The interneuronal components, characterized by the set of input-output filters $\{k_n\}$, can be intuited as the waveform of the EPSP from the $n^{th}$ input neuron onto the output neuron. A nonlinear filter is potentially included to describe interactions between two input pulses, such as paired pulse facilitation and depression (Song, Marmarelis, and Berger, 2009; Sandler et al., 2015b). More detailed information on model structure can be found in the appendix. It should be noted that the model is a "blackbox", or entirely based on data, and thus makes no *a priori* assumptions on the nature of the feedback and interneuronal dynamics. Therefore, the estimated filters include the previously listed phenomena as well as more indirect/nonlinear phenomena such as dendritic integration, spike generation, active membrane conductances, and feedforward interneuronal inhibition (thereby allowing the interactions between two pyramidal cells to be inhibitory).

All GLM parameters, which implicitly describe both connectivity and causal dynamics, were fit simultaneously for each neuron using MCP group regularization and a coordinate descent algorithm (see appendix 4.3.2). It should be noted that because parameters were estimated from spontaneous data rather than from direct perturbations of the network, all parameter estimates may be biased by unobserved inputs (see discussion). A Monte Carlo style shuffling approach was used to insure all obtained models had significant predictive power and were not simply overfitting (see appendix 4.3.3).

## 2.3 Simulation & Clustering

After neuronal connectivity is estimated, network activity can be simulated in a forward manner using the estimated coefficients. For each time bin $t$, the output of the $n^{th}$ neuron, $\tilde{y}_n(t)$ is modeled by an inhomogeneous Bernoulli process (i.e. biased coin flip) whose probability of firing is determined by the past and present spiking activity of other connected neurons:

$$\tilde{y}_n(t) = \begin{cases} 1 & \Phi(\tilde{\eta}_n(t), \sigma) > u \\ 0 & \Phi(\tilde{\eta}_n(t), \sigma) \leq u \end{cases} \quad (1)$$

Here $\Phi()$ is the probit link function (Eq. A2), $\tilde{\eta}_n(t)$ is the linearly weighted combination of past and present spiking activity for neuron $n$ (Eq. A3), and $u$ is a standard uniform random number. Thus the neuron spikes when the generated randon number, $u$, is greater than the neuron's instantaneous probability of firing, $\Phi(\tilde{\eta}_n(t), \sigma)$. Equivalently, the neuronal output can be viewed as being generated by a prethreshold signal, composed of the sum of a deterministic component (Eq. A4) and Gaussian white noise (GWN) of variance $\sigma^2$, followed by a fixed threshold of 0 which is implicit in probit link formulation (Berger et al., 2012):

$$\tilde{y}_n(t) = \begin{cases} 1 & \tilde{\eta}_n(t) + \mathcal{N}(0, \sigma^2) > 0 \\ 0 & \tilde{\eta}_n(t) + \mathcal{N}(0, \sigma^2) \leq 0 \end{cases} \quad (2)$$

Note that due to the stochastic nature of the simulation and the neuronal baseline firing rates, which are determined by the $k_0$ coefficients (see appendix), the model will generate spiking activity spontaneously, even without any explicit exogenous input.

It was found that when the RNN simulation was allowed to run, it would spontaneously enter distinct network states for extended time periods. A clustering algorithm was used to identify the distinct stable dynamical states which emerged in the RNN output (Sasaki, Matsuki, and Ikegaya, 2007; Burns et al., 2014). To apply the clustering algorithm, the instantaneous mean firing rate (MFR) was first computed for each neuron using a 100 ms moving average filter. 5 principal components (PCs) were then extracted from the 24 instantaneous MFR vectors (Fig. 3c). These 5 PCs were then clustered using a standard k-means algorithm. Finally, for each state, the dominant subnetwork of significant neurons was found by identifying the smallest set of neurons which together accounted for $> 85\%$ of the cluster energy (see Fig. 3e).

## 2.4 Stimulation Optimization

The RNN framework allows binary external stimulation, $s(n,t)$ to be applied to neuron (or 'electrode') $n$ at time $t$ by superimposing a spike onto that neuron at that time. Essentially, Eq. 1 is modified to be:

$$\tilde{y}_n(t) = \begin{cases} 1 & \Phi(\eta_n(t), \sigma) > u \text{ or } s(n,t) = 1 \\ 0 & \Phi(\eta_n(t), \sigma) \leq u \end{cases} \quad (3)$$

Intuitively, this model of stimulation allows electrodes to deterministically elicit spikes at precise times and thereby 'override' the neurons' normal probabilistic mode of spiking defined in Eq. 1.

After seizure and non-seizure states were identified, a two-stage simulated annealing (SA) algorithm was used to identify the optimal spatiotemporal stimulation pattern, $s^*(n,t)$, which would most quickly and reliably move the network from the seizure state to the non-seizure state (Kirkpatrick and Vecchi, 1983). Multiple SA streams were run in parallel with 2 different modes of stimulation: periodic spiketrains (PTs) and random (Poisson) spiketrains (RTs). Each electrode could stimulate at 8 possible parameter values: {OFF, 5 Hz, 20 Hz, 60 Hz, 100 Hz, 140 Hz, 180 Hz, 220 Hz}. The electrode parameter determined the periodic stimulation frequency for PTs and the Poisson rate for RTs. Note that while for PTs, each parameter vector determines a unique spatiotemporal pattern, for RTs, each parameter vector only specifies the firing probability, and could correspond to a near infinite amount of different spatiotemporal patterns. To further reduce the amount of parameters, the electrodes for neurons which were functionally disconnected from the RNN (i.e. had no inputs or outputs) were kept off.

The SA algorithm was run on a standard exponential cooling schedule with 120 global iterations with temperatures logarithmically spaced between $T_{start} = 10$ and $T_{end} = .01$. Each global iteration had 80 local iterations where the temperature was kept constant. At the end of each global iteration, the parameter was reset to the minimum of the global iteration (Henderson, Jacobson, and Johnson, 2003).

In every SA iteration, the network was initialized in the seizure state. The neurostimulation pattern, $s(n,t)$, was designed based on the current parameter values, and applied for 250ms. Since the seizure state was most obviously associated with high firing rates, the SA algorithm aimed to attain the stimulation pattern which most lowered the network MFR in the 100 ms after stimulation ended. Thus the cost function to be minimized was the ratio of stimulus-induced MFR to reference MFR (i.e. the MFR when no stimulation was applied). Note that the lowest value the cost function could take was 0, and when it was $> 1$, the stimulation actually intensified the seizure. At each iteration the same stimulus was applied 3 times, with different random number seeds, and the average MFR ratio over the 3 runs was used as the SA cost function. A sample optimization path can be seen in Fig. 4a.

The SA algorithm was run in two consecutive rounds. In both rounds one electrode was chosen whose stimulation parameter value would randomly transition to a neighboring value with equal probability; however, the definition of neighboring values was different for both rounds. In the first round, the neighboring values were defined as the immediately higher frequency, the immediately lower frequency, and OFF. If the selected electrode was already OFF, all 7 of the frequency values were defined as neighbors (i.e. an OFF electrode would transition to a random frequency with equal probability). Allowing an electrode to turn off at each iteration encouraged sparsity and facilitated the identification of a subset of electrodes to more carefully optimize in the second round. In the second round, the OFF electrodes from the first round were not adjusted and the remaining electrodes transitioned either to the higher or lower frequency with equal probability.

As the SA algorithm does not explicitly penalize nonsparse solutions, a stepwise pruning algorithm was used to remove superfluous electrodes. During each cycle, the algorithm individually iterated though all $E$ "on" electrodes and calculated the cost function if the selected electrode was "turned off". At the end of each cycle the electrode which caused the greatest decrease in the cost function was turned off, and the algorithm would continue to the next cycle and consider only the $E-1$ remaining electrodes. If no electrodes were found whose removal decreased the cost function, then the pruning algorithm stopped and no further electrodes were turned off.

## 3 Results

### 3.1 Reconstructed Neuronal Network

This study aims to design a realistic testbed for developing and screening efficient neurostimulation patterns for seizure abatement. This testbed, which we have dubbed a reconstructed neuronal network (RNN), is estimated from single unit activity in area CA3 of a human TLE patient undergoing monitoring for resective surgery (see methods). $R = 24$ distinct units were identified and used for the RNN.

The obtained connectivity graph is shown in Fig. 1a. The optimization regularization path is shown in Fig. 2a,b. Further metrics quantifying the predictive power of the estimated models, including ROC plots and the KS-test are shown in Fig. 2c,d. 18 out of 24 neuronal models and 22.8% of all possible connections were found to be significant. The remaining 6 neurons were functionally isolated from the network: they neither influenced any other neurons or were influenced by them. Of the significant connections, a much larger number than expected were found to be bidirectionally connected (71.43%, $P < .001$. Fig. 1b). All neurons had an average of 5.3 inputs and outputs. Furthermore, there was a positive correlation between number of inputs and number of outputs, suggesting a small-world network (Fig. 1c; Bullmore and Sporns (2009) and Fallani et al. (2014)).

A sample system is shown in Fig. 1d for neuron 22 which is causally influenced by neurons $\{2, 8, 14\}$. Several features of the system can be interpreted from the filters. Note that neurons 8 and 14 are connected linearly while neuron 2 is connected with a quadratic filter, thus implying it exerts some form of short term potentiation. Also note that neuron 14 exerts an entirely excitatory influence on neuron 22, while the effect of neuron 8 oscillates between excitation and inhibition. Finally, the feedback kernel is composed of initial refractory inhibition, followed by oscillatory bursting activity. Interestingly it oscillates at 5 Hz, which is in the low theta range. This oscillation has been extensively implicated in memory tasks in the hippocampus (Buzsaki, 2006; Sandler et al., 2014).

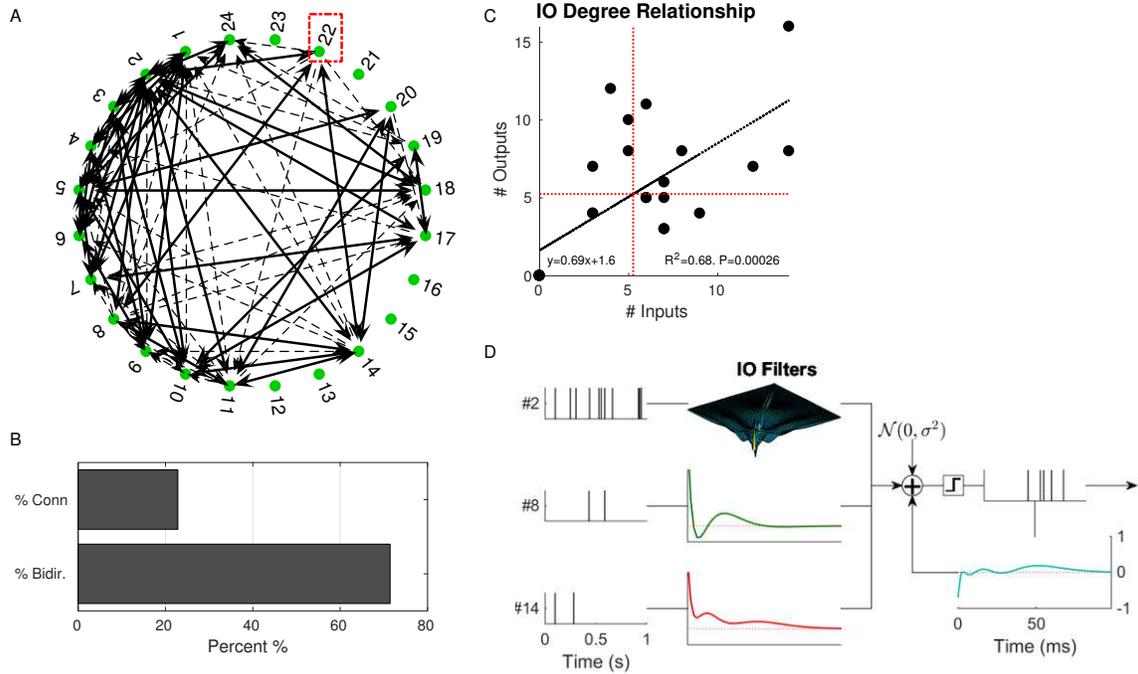

Figure 1: (A) Graph of all identified effective connections. Dashed lines indicate unidirectional connections, while solid lines indicate bidirectional connections. Note that some neurons (like 23) are effectively disconnected from the population. (B) Barplot showing % of total possible connections and the % of those which are bidirectional. (C) Positive correlation of the # of outputs a given neuron has vs the # of inputs it has. Regression data is given in the figure for best fit line. Red lines shown means of x and y data. (D) Sample three input system of neuron #22 (enclosed in red box in A). Input spiketrains are convolved with either a linear or quadratic filter and then summed with Gaussian noise and feedback effects to generate the prethreshold sum. They are then put through a threshold of 0 to generate a binary output.

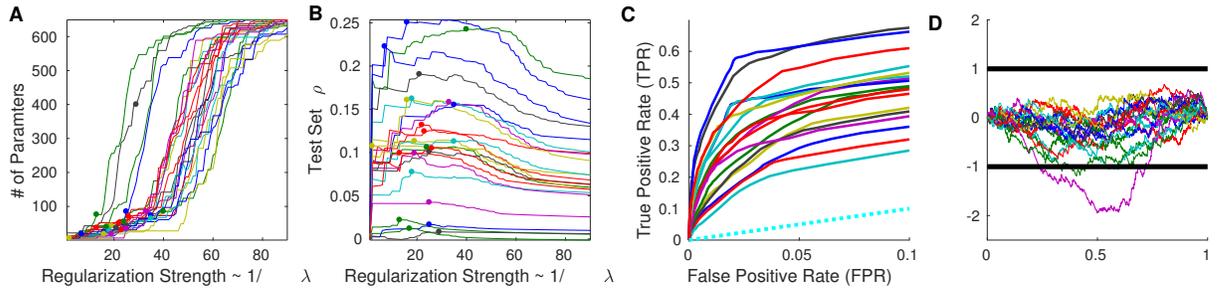

Figure 2: (A,B) Plots show # of parameters selected (A) and testing set correlation, $\rho$, (B) for different values of the regularization parameter, $\lambda$. Each plot shows a line for each of the 24 neurons in the RNN. Notice that as in $1/\lambda$ is increased (and thus regularization is weakened), more parameters enter the model until we have a full model. Dots show the optimal $\lambda$ selected for each neuron. (C) ROC plots for each of the 24 neurons, showing model predictive power. Notice that each of the lines are above the dashed blue line (TPR=FPR) which represents a model with no predictive power. (D) vertical KS Plots for each of the 24 neurons, along with normalized 95% confidence bounds (Song et al., 2013). As can be seen, most models fall within the bounds.

## 3.2 Seizure Initiation and Classification

Once the effective connectivity and dynamics were estimated, the RNN was allowed to run without perturbation and stochastically generate simulated hippocampal CA3 activity, $\{\tilde{y}_n(t)\}$ for all 24 neurons (Pillow et al., 2008). As expected the RNN, which was estimated from normal (nonictal) spiking activity of a TLE patient, generated physiological firing rates with very low levels of synchronization (Fig. 3a). In order to induce seizure dynamics, two modification were performed. First, the 'in-silico' neuronal membrane potential was raised by increasing the baseline firing probability parameter $k_0$ (see Eq. A3). This mimics the common practice of inducing seizures experimentally by pharmacologically raising the membrane baseline potential (Fricker, Verheugen, and Miles, 1999; Avoli et al., 2002). Second, the level of stochastic noise driving the network was reduced by lowering $\sigma$. To intuitively understand this modification, note that in the extreme case of $\sigma = 0$, the network is entirely deterministic and generates either no activity at all oscillates in a fixed limit cycle; alternatively, in the other extreme of $\sigma = \infty$, the network generates completely random (Poisson) firing. Thus, intuitively, lowering sigma increases population control over the neurons and tends to promote the persistent oscillations which characterize seizures. It was found that raising the baseline by $B = 30\%$ relative to the threshold and lowering $\sigma$ to .725 was sufficient to generate spontaneously emerging realistic seizures lasting anywhere between a few seconds to over a minute as seen in real human data (Fig. 3b; Bower et al. (2012) and Truccolo et al. (2014)).

In order to gain more intuition about the network dynamics, principal components (PCs) were extracted from the network activity based on instantaneous MFR. Fig. 3c shows the network trajectory through time in PC space. A clustering algorithm was used to identify the distinct stable dynamical states which emerged in the RNN spiking activity (Fig. 3d-f; Sasaki, Matsuki, and Ikegaya (2007) and Burns et al. (2014)). As can be seen, under the selected $\{\sigma, B\}$ parameters, the RNN jumped between only 2 stable states: normal and seizure. The dominant subnetwork of neurons which comprised the seizure cluster is shown in Fig. 3e. These 6 neurons are responsible for most of the spiking activity within the seizure state, and their reciprocal connectivity is presumably responsible for maintaining the seizure dynamics. Many neurons maintained their regular firing rate, and 2 neurons even reduced their firing rate. These observations match the heterogeneity of neurons in recorded human seizures (Bower et al., 2012). Finally, the cluster state of the network was identified at each moment in time (Fig. 3b,bottom). As can be seen, this method is able to reliably detect when the RNN enters and leaves the seizure state, thus making it a simple seizure detection algorithm (Mormann et al., 2007; Cook et al., 2013; Nagaraj et al., 2015).

## 3.3 Identifying Optimal Neurostimulation for Seizure Abatement

Once the effective connectivity of the human TLE patient was estimated, and realistic seizure activity simulated, we aimed to identify a spatiotemporal pattern of electric stimulation which could reliably and efficiently induce the network to leave the seizure state. At first glance, this is a paradoxical task: we want to lower network spiking activity by applying external spikes to the network. However, several experimental studies have shown that this could be accomplished using precisely designed patterned stimulation (Durand and Bikson, 2001; Heck et al., 2014).

Our working assumption was that there exist 24 electrodes which could each stimulate a single neuron without affecting any of the others. A "pulse" in an electrode at a given time would elicit a single contemporaneous spike in the associated neuron at that time. The experimental implications of these assumptions will be discussed later. Computationally, however, this introduces a highly complex optimization problem. If we only consider whether a given electrode will be on or off, there are $2^{24}$, or over 16 million possibilities. If any of the 24 could take on an arbitrary pattern of spikes over 250 ms, this number would increase to $2^{3000}$. To make the problem less formidable, only two modes of stimulation were considered: periodic trains (PTs) having a fixed frequency of stimulation, and random, or Poisson, trains (RTs) with different MFRs (see Fig. 5b). Thus, every electrode was considered a parameter which could take on 8 values, including $\{OFF, 5Hz, 20Hz, 60Hz, 100Hz, 140Hz, 180Hz, 220Hz\}$. This allowed a total of $2*8^{24}$ possible stimulation types.

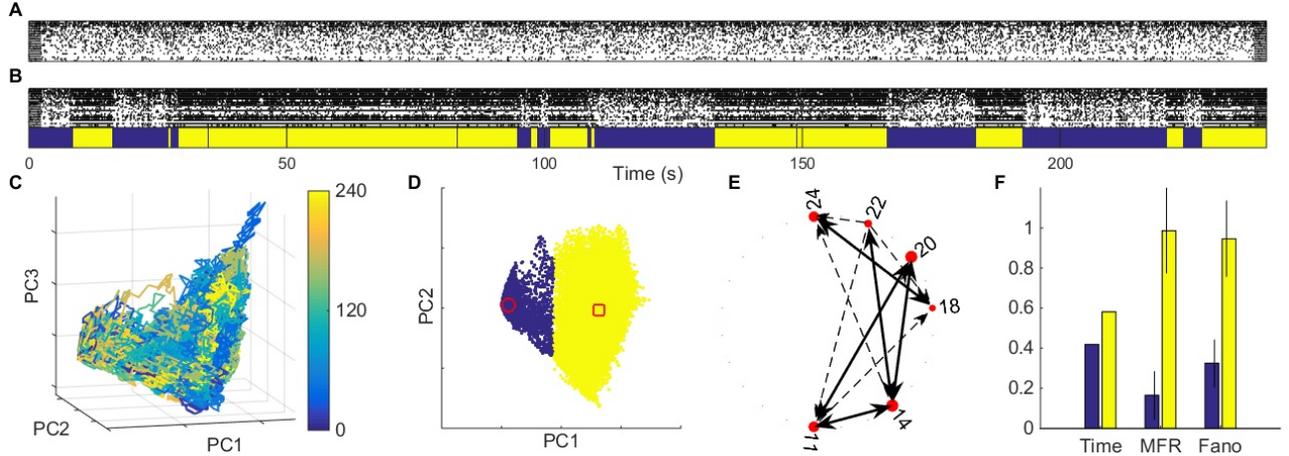

*Figure 3: (A) 4 minutes of simulated firing of RNN in physiological conditions. (B) Top: Simulated firing of RNN after modifications to induce seizures. Notice the RNN spontaneously enters and leaves the seizure state. Bottom: the network cluster state through time (see D). (C) Trajectory of RNN activity in (B) within the space of the first 3 principal components. Color indicates time. (D) Results of clustering algorithm applied to PC trajectory. Red dots indicate cluster centers. (E) The connectivity between the subnetwork of neurons which were found to dominate the seizure cluster (yellow). Bigger circles indicate bigger MFRs during seizures. (F) Additional metrics characterizing the two clusters, including proportion of time, MFR, and Fano factor in each state. Note that MFR was scaled to have a maximum of 1 to promote visualization.*

The stimulation length was fixed to 250 ms and a two-stage simulated annealing algorithm was used to find the optimal parameters which would make the network leave the seizure state most quickly (see methods). The path of the simulated annealing algorithm is shown in Fig. 4a. The optimal stimulation parameters and spatiotemporal pattern are shown in Fig. 4c,d. It was found that only 4/24 ( 17%) electrodes needed to be turned on. Of those, 2 electrodes had frequencies of 180 Hz, and the remaining two had frequencies of 100 Hz and 220 Hz; furthermore, it was found that periodic stimulation led to better results than Poisson stimulation (Fig. 4b). This confirms experimental evidence showing that high frequency stimulation (HFS) is optimal for abating seizures (Durand and Bikson, 2001). Interestingly, 2 of the selected electrodes (22 and 24) stimulated the epileptic subnetwork (Fig. 3e), while the other two stimulated outside the subnetwork, suggesting that direct stimulation of the seizure focus itself may not be the most effective route.

By initializing the RNN simulations under identical seizure conditions and using identical sequences of random numbers, one could compare how a seizure would have evolved under different types of applied stimulation. Fig. 4e,f show how a network initialized in the seizure state evolves when no stimulation is applied (reference) and when the optimal stimulation in Fig. 4d is applied. Additionally, Fig. 4g,h shows the population MFR and PC trajectory in both simulations. As can be seen from these figures, the optimal stimulation is able induce the network to leave the seizure state in under 250 ms, and more importantly the network stays in the nonictal state after the 250 ms stimulation ended.

Our working premise has assumed that precisely designed independent, or unsynchronized, stimulation across multiple sites could improve responsive neurostimulation. While some work has supported this hypothesis (Nelson et al., 2011), most DBS studies, due to either experimental or theoretical consideration (Durand and Bikson, 2001), have only looked at single site stimulation where an electric pulse presumably stimulates a large population of neurons simultaneously (Sun and Morrell, 2014). Furthermore, most studies have used periodic spiketrains (PTs) rather than random spiketrains (RTs) despite a few studies indicating that RTs may be superior to PTs (Fig. 5a, Wyckhuys et al. (2010) and Van Nieuwenhuyse et al. (2014)). In order to compare synchronized PT and RT stimulation with independent multi-electrode stimulation we first attempted to find the optimal stimulation frequency/rate for PT/RT stimulation. This was done by delivering identical patterns of stimulation to all 24 neurons simultaneously for 250 ms. The optimal frequency was found by sweeping from 5 Hz to 220 Hz in 50 Monte-Carlo style trials. Once again, in each trial, the RNN was initialized in the seizure state, and identical random numbers were used for each frequency of stimulation in order to allow a fair comparison of the stimulation frequencies under equivalent conditions (which notably is impossible in real life). The results are shown in Fig. 5b. Neither synchronized PTs or synchronized RTs, of any frequency, were found to significantly help in ending seizures. In fact, at many frequencies they actually exacerbated the length of seizures. Interestingly however, PTs and RTs at high frequencies ($> 180$ Hz) did temporarily move the network away from the seizure zone (Fig. 6). However, as soon as the stimulation was turned off, the seizure continued. It should be emphasized however, that these results cannot be generalized beyond the particular patient from whom this data was estimated, and high frequency stimulation has been shown to be efficacious in a large number of patients (Heck et al., 2014).

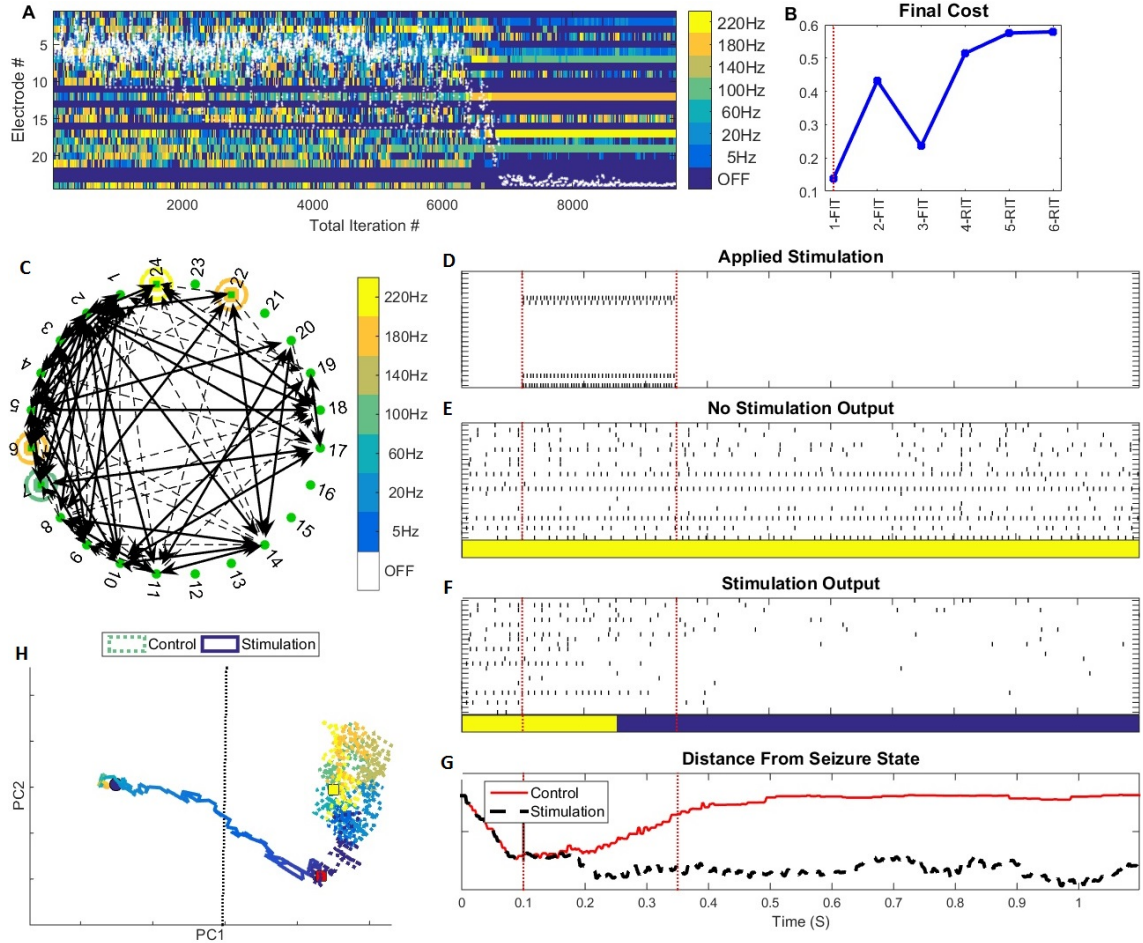

*Figure 4: (A) Evolution of electrode parameters during the second round of simulated annealing. Notice that some electrodes are consistently kept off as they were not selected during the first round. White dots show the relative cost function value for each iteration. (B) Final costs for 6 parallel runs of the SA algorithm using different modes of stimulation: PT,RNB, and RT. Vertical red line shows that the best results were achieved for PT stimulation and these parameters were the selected ones. (C) Directed connectivity graph from Fig. 1a, where the optimal electrode frequencies are indicated by the color of the circles surrounding the neurons. No circles indicates that neuron is not to be stimulated. (D) Rasterplot of optimal spatiotemporal pattern from the SA algorithm. Rasterplots of evolution of a seizure when no stimulation is applied is shown in (E), and when optimal stimulation is applied in (F). Cluster state is shown below both rasterplots (yellow=seizure, blue=normal). (G) Distance from the seizure state cluster center is shown for both reference and stimulation runs. Notice that stimulation induces the network to rapidly move away from the seizure cluster center. This can be seen more clearly in (H) which shows the trajectory of both runs in PC space (see Fig. 3c. Dotted red lines in (D-G) indicate the beginning and end of stimulation.*

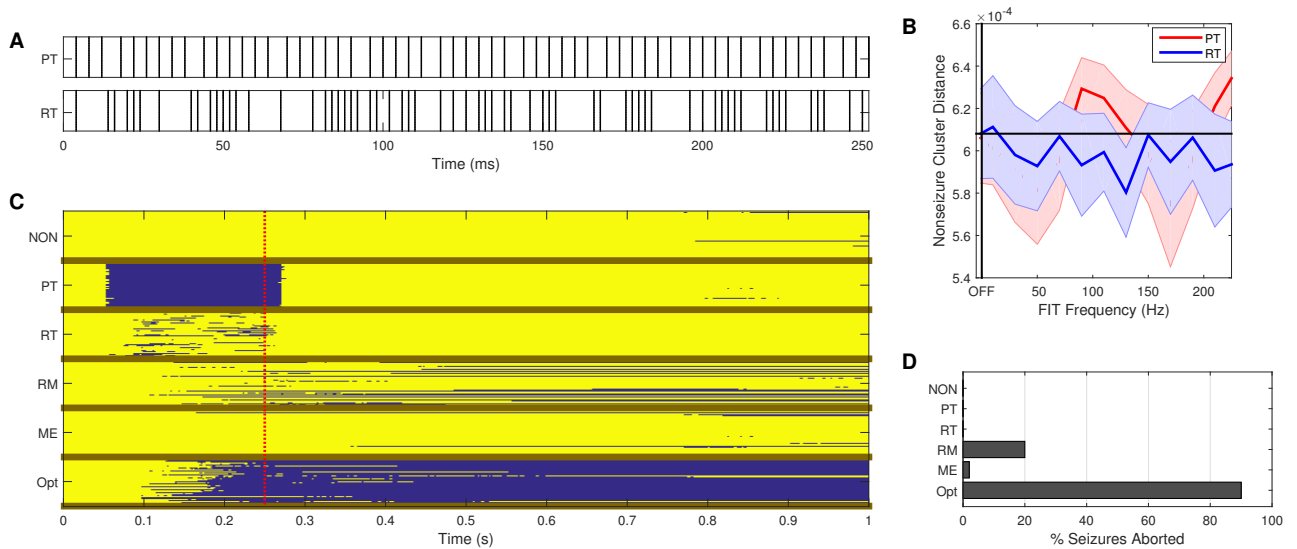

Figure 5: (A) Examples of optimal 170Hz PT (top) and 130Hz RT applied equally to all 24 neurons. (B) The optimal PT and RT frequencies were found by comparing seizure length over various frequencies over 50 trials. Black lines indicate seizure length in reference (no stimulation) conditions. Line and shading show mean±SEM. (C) Comparison of performance of 6 types of stimulation (see text) over 50 trials. Stimulation was applied for the first 500 ms (vertical red line). Each row shows the network cluster over time (see Fig.2b,d). As can be seen, only the optimal SA stimulation pattern was able to move the network from the seizure cluster (yellow) to the normal cluster (blue). (D) % of seizures aborted by various stimulation modalities.

Finally, the optimal stimulation was compared with other unsynchronized multi-electrode stimulation patterns to insure that the acquired simulated annealing solution is indeed a significant local, if not global, minima. In each simulation, the optimal stimulation was applied along with no stimulation (NON), and various control stimulation patterns including: (1) 200-Hz PT, (2) 200-Hz RT (Fig. 5a,b), (3) random unsynchronized multi-electrode (RM) stimulation having the same amount of selected electrodes and total bursts as the SA solution, and (4) stimulation using the same electrodes as the SA solution, but with mixed frequencies (MF). The results are shown in Fig 5c,d. Again, it can be seen that both types of synchronized stimulation failed to provide good results. Once again, it can be seen that synchronized stimulation can only temporarily move the network out of the seizure state. The mixed-frequency stimulus (MF) was only able to abate 5% of seizures confirming that the optimized HFS settings were needed for successful abatement. The random unsynchronized (RM) stimulation, which applied the HFS to random electrodes, was able to abate 20% of seizures showing that even random unsynchronized stimulation is, at least in this patient, superior to optimized synchronized stimulation. Finally, the optimal SA solution was able to abort 92% of the seizures. In the four unsuccessful cases, it was found that while the stimulation did push the network temporarily out of the seizure state, the seizure restarted after the stimulation was turned off. These results confirm the efficacy of the SA algorithm to identify optimal stimulation patterns.

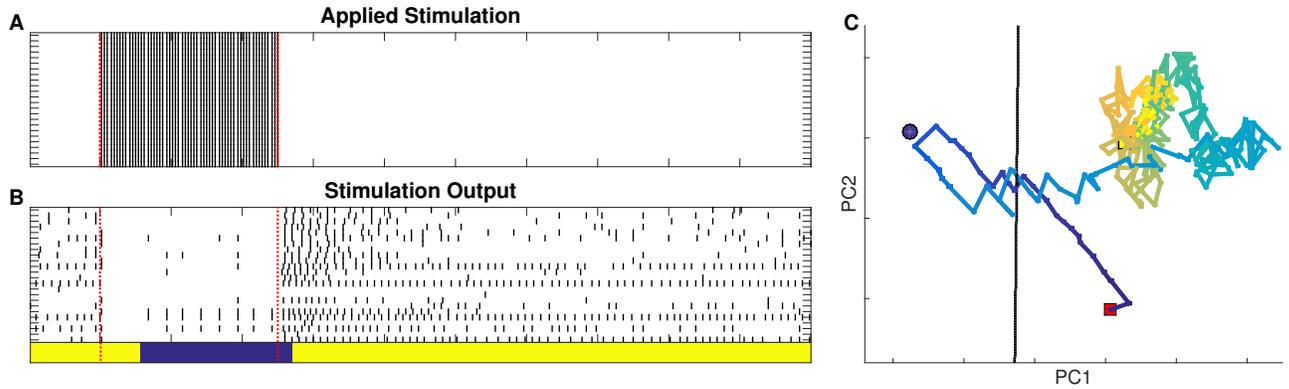

*Figure 6: (A) Synchronized 200 Hz periodic stimulus. (B) Network output. Notice that the seizures temporarily stops during stimulation; however, the seizure returns almost immediately once stimulation is turned off. (C) The same plot in principal-component space.*

### 3.4 Responsive Neurostimulation

In order to assess the feasibility of model-based responsive neurostimulation, a basic control strategy was used (Chakravarthy et al., 2007; Chakravarthy et al., 2009b; Schiff, 2012). The strategy consisted of applying the optimal 250 ms neurostimulation pattern (Fig. 4d) in 'real-time' as soon as a seizure was detected. Causal clustering allowed real-time seizure detection by identifying when the network shifted from the normal to seizure cluster (Fig. 3). At the end of the stimulation, if the network was still in the seizure state, another round of stimulation was immediately applied; otherwise, the stimulation therapy ended. Notably, this is the same control algorithm currently used in the Neuropace RNS® device (NeuroPace, Inc, 2015). (However, it should be noted that the Neuropace broadens our definition of 'seizure state' to include interictal activity, with the hope that stimulation will prevent the network from ever entering a full seizure.)

Two simulations were performed with and without responsive neurostimulation (Fig. 7). Both simulations were under epileptic $\{\sigma, B\}$ parameters and were performed using identical initial conditions and random number generators. As can be seen, in the reference (non-stimulation) case, 4 spontaneous seizures emerges lasting between 5 and 34 seconds. The responsive neurostimulation algorithm was able to detect all these seizures (and the additional seizures which would have emerged had the network not already been in the seizure state). Furthermore, in most cases a single round of stimulation was able to prevent the network from going into a prolonged seizure state. In one case, at 72 seconds, the first stimulus was unsuccessful eliminating the seizure, so a second stimulus was applied thereafter and successfully stopped the seizure.

## 4 Discussion

In this study single neuron activity from human hippocampus was used to develop a reconstructed neuronal network (RNN) which replicates in-silico the distinctive connectivity and causal dynamics of the recorded 24 neurons (Fig. 1). the RNN was estimated using a non-parametric/phenomenological approach based entirely on recorded data and which makes few *a priori* assumptions about the biophysical nature of the network dynamics (Pillow et al., 2008).

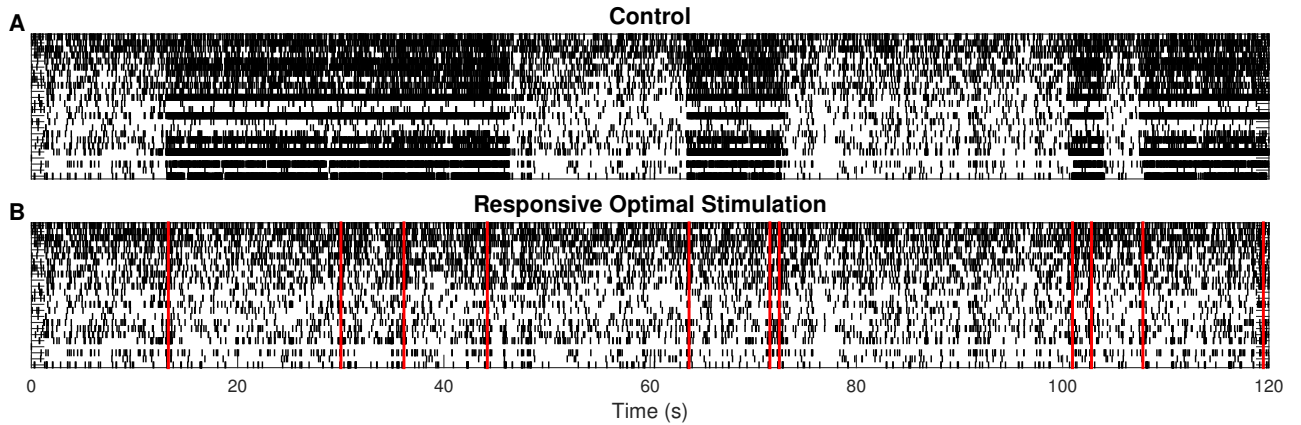

*Figure 7: (A) 2 minutes of spontaneous network activity under epileptic conditions where 4 seizures spontaneously emerged. (B) identical network activity in (A), but with responsive neurostimulation, whereby the optimal 250 stimulus was delivered at the times indicated by the red lines. As can be seen, the stimulation was able to avert a prolonged seizure.*

The spiking probability of each neuron was estimated using a realistic model incorporating the output neuron's past spiking history and the spiking history of all connected neurons. Group regularization allowed for the efficient and compact estimation of connectivity and model complexity (i.e. whether neuronal interactions are best described by a linear or nonlinear filter).

In this study recorded data from a human hippocampus was used to estimate the RNN. In principle the RNN could have been estimated from any simultaneously observed set of point processes, including those produced by artificial spiking neural networks. Human data was chosen because it most accurately reflects the conditions which would be met by such an algorithm in practice. Namely, it already inherently contains the desired level of complexity and avoids the need to make any assumptions on the stochasticity, connectivity, and higher order statistics of clinical data. Nonetheless, due to (1) the difficulty of obtaining human hippocampal single unit data and (2) the emphasis of this work on methodology and proof-of-concept only one dataset was used in the study. A central goal of future work will be to see how well the algorithm performs on a diverse set of synthetic and recorded datasets and to establish theoretical criteria for its success.

## 4.1 RNN for Neurostimulation Design

After the RNN was estimated, seizure dynamics were induced by raising membrane potential and isolating the network from external noise (Fig. 1), both features which have been implicated in initiating physiological seizures (Fricker, Verheugen, and Miles, 1999; Wendling et al., 2003; Warren et al., 2010). Finally, a simulated annealing algorithm was used to design an optimal stimulation pattern to induce the network to leave the seizure state (Fig. 4). The optimal stimulation was found to abate 92% of seizures and was successfully used in a responsive stimulation paradigm to prevent seizures from developing in the RNN (Fig. 5-7). These results lead us to hypothesize that (1) the unique nature of every patient's seizure focus proscribes any single neurostimulation pattern from being optimal in every patient (2) the distinctive nature of the seizure focus can be exploited to algorithmically develop efficient patient-specific neurostimulation patterns.

A conceptually similar approach has been used in a previous study where a neural mass model of the thalamocortical network estimated from patient data was used to explain why particular frequencies of stimulation were successful to abate seizures while others were not (Mina et al., 2013). Additionally, such a customized/algorithmic approach has already begun to be applied to develop neurostimulation patterns for Parkinson's Disease (PD) (Holt and Netoff, 2014; Grill, Brocker, and Kent, 2014; Brocker et al., 2017; Peña et al., 2017). However, when dealing with PD neurostimulation, one has near instantaneous feedback of the stimulation by assessing its effects on the patient's tremor (Brocker et al., 2013). The challenges are much greater in epilepsy where physicians must oftentimes wait several months before they can access the quality of a particular stimulation design due to the infrequency of seizures. Furthermore, physicians cannot "go back in time" to see how a particular seizure would have evolved had a different stimulation pattern been applied or had no stimulation been applied at all. This is particularly important since responsive neurostimulation aims to perturb the network in the preictal state and thus prevent the seizure from ever occurring; currently, however, devices such as the Neuropace suffer from a very high false-positive rate. This means that the lack of a seizure following stimulation cannot be used as indicative of its success since in most cases no seizure would have developed regardless of the stimulation pattern. Due to these difficulties, the Neuropace manual recommends that physicians use a 200 Hz periodic stimulus, and if it is unsuccessful to increase the current amplitude (NeuroPace, Inc, 2015). This is despite the fact that the device allows two leads to be independently programmed with a wide range of complex stimuli and a frequency range of 1-333 Hz (Sun, Morrell, and Wharen Jr, 2008). We believe a major bottleneck in the performance of devices such as the Neuropace, which currently reduces seizures by an impressive but far from perfect 54% (Heck et al., 2014), is not the hardware, but rather the physicians inability to successfully identify optimal stimulation parameters.

In this study, model based in-silico neurostimulation optimization is presented as a solution to this vital issue. In this paradigm, a patient-specific model is used as a testbed or hypothesis engine for the design and validation of optimal neurostimulation. This paradigm provides solutions to many of the experimental issues of neurostimulation design: one may obtain seizures "on-demand" by initiating the network in the seizure state. Furthermore, by using identical sequences of random numbers, one can "go back in time" and observe how the seizure would have evolved under different applied stimuli. Using this testbed we gained many insights into neurostimulation and generated many testable predictions. It was observed that synchronized random (Poisson) stimulation provides slight benefits over periodic stimulation - an observation previously observed in the literature both experimentally and in computational models (Wyckhuys et al., 2010; Buffel et al., 2014). However, neither of these stimulus styles performed as well as random unsynchronized stimulation over multiple sites, suggesting the need to conduct more experiments exploiting multiple electrodes for stimulation (Cook et al., 2013; Van Nieuwenhuyse et al., 2014). Furthermore, it was observed that our proposed optimal stimulation disproportionably targeted neurons outside of the epileptic subnetwork (i.e. focus). However, our most important finding was that optimized stimulation over multiple electrodes significantly outperformed any of the previously mentioned stimulus styles by exploiting the unique connectivity and dynamics of the RNN. Most importantly the optimal neurostimulation pattern identified here can be experimentally validated by applying it to the patient for which it was estimated on. This ability to experimentally validate our model predictions is lacking in many of the computational studies exploring neurostimulation.

## 4.2 Modeling Methodology and Limitations

While the current framework provided very strong computational results, several limitations need to be addressed before it can be applied experimentally. The network dynamics were estimated from spontaneous/observed data and predictive power was used to determine connectivity. This Granger-causality approach is biased by unobserved inputs. In this case, neurons within CA3 and neurons which project to CA3 (such as those from the entorhinal cortex) are potential unobserved inputs. Furthermore, our model of electrical stimulation where an electrode consistently elicits a single spike uniquely in the neuron that it records, is overly simplistic. Any stimulation will affect at least dozens of surrounding cells (Wei and Grill, 2005; Desai et al., 2014) and have complex temporal dynamics. Additionally, there is evidence that the neurons an electrode records and those which it stimulates are not equivalent (Histed, Bonin, and Reid, 2009). Both of these issues can simultaneously be overcome by experimentally perturbing the network using sequential stimulus pulses across multiple electrodes and analyzing the response. Efficient algorithms are already being developed for how to optimally design such experiments (Lepage, Ching, and Kramer, 2013; Kim et al., 2014)

The current model assumes stationarity of neuronal dynamics, while in reality recorded neurons exhibit strong nonstationarities due to synaptic plasticity and electrode drift. The stationarity assumption was necessary in order to define a tractable optimization problem. It is unknown to what extent nonstationarities in neuronal dynamics lead to nonstationarities in optimal stimulation parameters. Future work may aim to address this issue by attempting to identify optimal neurostimulation patterns in nonstationary networks such as those modeled in Robinson, Berger, and Song (2016).

The model presented here relied on single unit activity. In practice, however, recording and sorting stable single units is quite difficult (Gilja et al., 2012; Dhawale et al., 2017). Thus, in future work we shall apply the presented framework for identifying optimal stimulation to continuous electrophysiological signals such as ECoG. In this case, each of the specific steps would be modified, while the overall framework would remain the same. For example, in the simulated annealing algorithm the current MFR based cost function would need to be substituted for one which can be applied to continuous signals, such as one based on high frequency oscillations. Furthermore, due to the difficulty of recording single units during human seizures, and the difficulty of estimating reliable models from such short data records, the present work synthetically induced a seizure. Synthetic seizure dynamics were homogenous, while there is evidence that actual seizures exhibit diversity even within the same patient (Bower et al., 2012; Aarabi and He, 2014). In future work, estimating network dynamics directly from recorded seizure data will lead to more conclusive results as less assumptions about seizure initiation will need to be made. Most importantly, there is a need to validate the obtained results experimentally. We imagine that the advocated framework will need to go through several iterations of experimental refinement until the strong computational results achieved here can be matched in actual animal models or human patients.

From a computational perspective, several improvements can be made to the stimulation design and control algorithm. While the simulated annealing algorithm considered a very large space of stimulation possibilities ($2^{176}$ total), several stimulation patterns were not considered such as those which intermixed periodic and random pulse trains and completely arbitrary pulse trains (Grill Jr and Brocker, 2014; Brocker et al., 2017). Furthermore, results may potentially be improved by incorporating the relative phases between different pulse trains into the algorithm. Also, a relatively simple control strategy based on the Neuropace device was employed. The advantage of this strategy was that the stimulation was independent of the seizure specifics. In the future more sophisticated control algorithms may be employed which emit different stimulation patterns based on the quality and progression of the specific seizure (Ching, Brown, and Kramer, 2012; Kalitzin et al., 2014; Ehrens, Sritharan, and Sarma, 2015; Tsakalis and Iasemidis, 2006; Tsakalis et al., 2006; Jassemidis and Tsakalis, 2012; Schiff, 2012). Overall, while many improvements can be made in the specifics, we believe that the overall framework presented here has the potential to significantly increase the effectiveness of neurostimulation for epilepsy.

## 4.3 Vision

Our speculative and perhaps overly optimistic vision is that in the future epileptic patients will be implanted with stimulation devices consisting of multiple electrodes which are capable of both recording and stimulating (Ryapolova-Webb et al., 2014) and can be independently programmed. Upon implantation, an automatic stimulation algorithm will perturb the network to establish safe current levels and to map effective connectivity between the observed areas. Machine learning algorithms will then program the initial stimulation parameters. As is currently done in the Neuropace (NeuroPace, Inc, 2015), the device will automatically record all detected epileptic activity and this data will be uploaded daily to a computer. Then, this data and patient input will be used offline to analyze the success of yesterday's stimulation. Finally, a reinforcement learning paradigm (Gosavi, 2014), will be used to adjust parameters for the next day. While admittedly, such a task may seem incredibly difficult to realize we believe that the growth of machine learning in the last decade has made this more realistic to accomplish than ever before. Furthermore, since this approach is fundamentally an algorithmic one and does not heavily rely on specific electrode design and placement, it may be backwards compatible with current generation devices and significantly improve their efficacy.

## Appendix: Dynamic Connectivity Model and Estimation

### 4.3.1 Model Structure

A probabilistic model was used to predict the firing probability of a given output CA3 neuron based on its own spiking history and the past and present spiking activity of all other functionally connected CA3 neurons. Thus the probability that a particular neuron, $y(t)$ will fire at discrete timebin $t$ is expressed by the probability, $\hat{y}(t)$:

$$\hat{y}(t) = Pr\Big(y(t) = 1 | x_1(t-\tau)...x_N(t-\tau), y(t-1-\tau)\Big) = H\Big[x_1(t-\tau)...x_N(t-\tau), y(t-1-\tau)\Big] \quad (A1)$$

where $\{x_n(t)\}$ reflect the $N$ effectively connected spiketrains from CA3, $\tau$ reflects the finite memory of the system which ranges from $0 \leq \tau \leq M$, and $H[\,]$ reflects the mathematical model which is used to describe the dynamical transformation from $\{x_n(t)\} \to y(t)$. The generalized linear modeling (GLM) framework was used whereby $H[\,]$ was decomposed into a linearized function of the inputs, $\eta(t)$, followed by a static nonlinearity, here chosen to be the probit link function (Truccolo et al., 2005; Song et al., 2007):

$$\hat{y}(t) = \Phi\Big(\eta(t), \sigma\Big) = \frac{1}{\sqrt{2\pi\sigma^2}} \int_{-\infty}^{x} e^{-\frac{1}{2}(\frac{\eta(t)}{\sigma})^2} \quad (A2)$$

$\eta(t)$, the linearized component of the GLM, takes the form of a nonparametric multiple-input autoregressive model which describes the dynamical transformation between input and output spike trains. It consists of a feedforward component, reflecting the effect of the $N$ input cells on the output cell, and a feedback/autoregressive component reflecting how the cell's past spiking history affects its current probability of spiking. Thus, the output is calculated as:

$$\eta(t) = \underbrace{k_0}_{\text{baseline}} + \underbrace{\sum_{n=1}^{N} F[x_n(t), k_n]}_{\text{interneuronal}} + \underbrace{F[y(t), k_{AR}]}_{\text{feedback}} \quad (A3)$$

where $F[x_n(t), k_n]$ models the feedforward effects of input $x_n(t)$, $F[y(t), k_{AR}]$ models feedback effects, and $k_0$, the constant offset term, models the baseline firing probability.

In past studies, feedforward effects were fixed to be either linear (Sandler et al., 2014) or nonlinear (Sandler et al., 2015b; Song et al., 2007), and feedback effects were fixed to be linear (Song et al., 2007). In this study, using the sparse group selection algorithm (see section 4.3.2), it is possible to determine whether particular inputs and feedback effects are best modeled by either a linear or nonlinear kernel (see Fig. 1d). In this study, linear refers to convolution with a linear filter, $k^{(1)}(\tau)$, while nonlinear refers to convolution with a quadratic (2nd order Volterra) filter, $k^{(2)}(\tau_1, \tau_2)$ (Marmarelis, 2004; Rajan, Marre, and Tkačik, 2013). Mathematically, these operations are respectively defined by Eq.A4a,b:

$$F[x(t), k^{(1)}(\tau)] = \sum_{\tau=0}^{M} k^{(1)}(\tau)x(t-\tau) \qquad (A4a)$$

$$F[x(t), k^{(2)}(\tau_1, \tau_2)] = \sum_{\tau_1=1}^{M}\sum_{\tau_2=1}^{M} k^{(2)}(\tau_1, \tau_2)x(t-\tau_1)x(t-\tau_2) \qquad (A4b)$$

It was found that a memory of 100 ms was sufficient to model the dynamical effects of most neurons, and thus $M$ was fixed to 50 (100 ms/2 ms binwidth). In order to reduce the amount of model parameters and thereby increase parameter stability, we applied the Laguerre expansion technique (LET) to expand the feedforward and feedback filters over $L$ Laguerre basis functions (Marmarelis, 2004). Based on previous studies, $L = 6$ Laguerre basis functions were used (Sandler et al., 2015a; Song et al., 2007; Song et al., 2009). Correspondingly, the amount of parameters in linear kernels was reduced from $M$ to $L$ (savings of 44 parameters) and in 2nd order kernels from to $M(M+1)/2$ to $L(L+1)/2$ (savings of 1254 parameters). The Laguerre parameter $\alpha$ was fixed at 0.542 to reflect this system memory (Marmarelis, 2004).

### 4.3.2 Model Estimation

Presumably, only a small portion of the total recorded neurons, $R$, causally influence any given neuron in the reconstructed neuronal network (RNN). This motivates the central task of identifying which neurons are effectively connected and which are not (Bullmore and Sporns, 2009; Fallani et al., 2014). Most methods which aim to estimate effective connectivity adopt a Granger causality approach whereby neuron A is effectively connected to neuron B only if it can help predict when neuron B will spike (Krumin and Shoham, 2010; Kim et al., 2011; Sandler et al., 2014; Zhou et al., 2014). Here a penalized group regression approach was adopted which implicitly maximizes predictive power while improving computational efficiency over commonly used stepwise input selection methods (Song et al., 2013; Song et al., 2016).

To proceed with penalized group regression, Eq. A2 is first recast in matrix form:

$$\hat{\boldsymbol{y}} = \Phi(\boldsymbol{\eta}) = \Phi(\boldsymbol{V}\boldsymbol{c}) \qquad (A5)$$

where $\hat{\boldsymbol{y}}$ and $\boldsymbol{\eta}$ are the vectors consisting of $\hat{y}(t)$ and $\eta(t)$ for $1 \leq t \leq T$, $\boldsymbol{V}$ is the design matrix consisting of the convolved inputs and, for quadratic kernels, their cross products (Marmarelis, 2004), and $\boldsymbol{c}$ is the vector of model parameters to be estimated. In the 24 neuron RNN, there are 300,000 observations and 649 unknown parameters.

The input parameters are divided into $2R$ groups consisting of 2 groups for every putative input: one group for the $L$ 1st order kernel parameters and another for the $L(L+1)/2$ 2nd order kernel parameters. Thus, for the $R = 24$ recorded neurons, there were 48 groups: 24 1st order kernel groups and 24 2nd order kernel groups. The objective is now to find the optimal parameter vector, $\boldsymbol{c}^*$ which minimizes the cost function composed of the sum of the negative log-loss likelihood and the group regularization term, $P(\boldsymbol{c})$:

$$C(\boldsymbol{c}; \boldsymbol{y}, \boldsymbol{V}) = \sum_{t=1}^{T} \Big( y(t) log\hat{y}(t) + (1-y(t))log(1-\hat{y}(t)) \Big) + \sum_{g=1}^{2R} P(\boldsymbol{c_g}) \qquad (A6)$$

where $\boldsymbol{c_g}$ is the group parameter vector containing only the parameters within group $g$. The function of the regularization term is to automatically set to 0 any parameter groups which are not found to significantly influence the output - thus implicitly estimating sparse functional connectivity. Here, the group minimax concave penalty (MCP) regularizer was chosen over the more conventional group LASSO because (1) it induces much less regularization based parameter shrinkages (biases) than the latter (2) it leads to much sparser solutions than the latter (Breheny and Huang, 2014; Zhang, 2010). The MCP regularizer is defined as:

$$P(\boldsymbol{c_g}; \lambda, \gamma) = \begin{cases} \lambda \|\boldsymbol{c_g}\| - \frac{\|\boldsymbol{c_g}\|^2}{2\gamma} & \|\boldsymbol{c_g}\| \leq \gamma \lambda \\ \frac{1}{2}\gamma \lambda^2 & \|\boldsymbol{c_g}\| > \gamma \lambda \end{cases} \quad (A7)$$

where $\lambda$ determines the strength of regularization and $\gamma$, which was fixed at 3, determines the range over which MCP regularization is applied (Breheny and Huang, 2014). The group coordinate descent algorithm outlined in Breheny and Huang (2014) was used to find $\boldsymbol{c}^*$.

The optimal $\lambda$ value was selected using a warm-start regularization path approach using 90 logarithmically spaced $\lambda$ values. At each iteration, the Pearson correlation, $\rho$ was computed between the recorded spiketrain, $y(t)$, and the estimated spike probabilities, $\hat{y}(t)$, on a testing set consisting of 20% of data randomly selected from $\boldsymbol{V}$. $\rho$ was used since (1) it was previously found to be a robust metric of similarity between spiketrains and continuous signals (Sandler et al., 2014) and (2) it led to sparser solutions than the more commonly used cross-entropy error. Finally, the optimal $\lambda$ was selected as the largest $\lambda$ which achieved $> 99\%$ of the max $\rho$ value. The regularization path can be seen in Fig. 2.

Since any regularization will necessarily bias obtained parameters to varying degrees, all parameters of nonsparse groups were reestimated without sparse groups and without the regularization term. Note that due to computational efficiency and simplicity the initial search for $\lambda$ uses a logit link function (Breheny and Huang, 2014), while the final reestimation was done using the probit link of Eq. A2. All computations were done in Matlab using custom code available upon request. A standard 3.2Ghz, 6-core desktop computer was able to estimate the 24-neuron RNN in approximately 3 hours.

### 4.3.3 Model Validation

To avoid overfitting, Monte Carlo style simulations were used to select those models which represent significant causal connections between input and output neurons and do not just fit noise (Sandler et al., 2014). The following procedure was used: in each run the real input was divided into 40 blocks and these blocks were randomly permuted with respect to the output. A model was then generated between the permuted inputs and the real output, and the Pearson correlation coefficient, $\rho_i$, was obtained as a metric of performance. $T = 40$ such simulations were conducted for each output and a set of performance metrics, $\{\rho_i\}_i^T$, was obtained. Then, using Fisher's transformation, we tested the hypothesis, $H_0$, that $\rho$ was within the population of $\{\rho_i\}$. If this hypothesis could be rejected at the 99.99% significance level, the model was deemed significant. The very conservative threshold ($P < .0001$) was used due to the large amount of comparisons being made.

Two metrics were used to evaluate the goodness-of-fit of the estimated models by comparing the estimated continuous output, $\hat{y}(t)$ with the true binary output $y(t)$. Receiver Operator characteristic (ROC) curves plot the true positive rate against the false positive rate over the putative range of threshold values for the continuous output, $y(t)$ (Fig. 2c, Zanos et al. (2008)). The second metric used was the discrete KS test (Haslinger, Pipa, and Brown, 2010; Song et al., 2013) which compares the ISI distribution of the time rescaled probabilistic estimates with that of a homogenous Poisson process (Fig. 2d). All model assessment metrics were evaluated on the testing set.

## Acknowledgements

This work was supported by NIH grant P41-EB001978 to the Biomedical Simulations Resource at the University of Southern California and DARPA contracts N6601-14-C-4016 and N66601-09-C-2081. The authors thank Dr. Daniel E. Couture, and Dr. Gautam Popli, Wake Forest Baptist Medical Center for their contribution of human hippocampal neural recordings.